\documentclass[twocolumn,prl,aps]{revtex4} 
\usepackage{amssymb}
\usepackage{epsfig,amsmath}
\usepackage{graphicx}% Include figure files
\usepackage{dcolumn}% Align table columns on decimal point
\usepackage{bm}

\newcommand{\beq}{\begin{equation}}
\newcommand{\eeq}{\end{equation}}
\newcommand{\beqa}{\begin{eqnarray}}
\newcommand{\eeqa}{\end{eqnarray}}

\newcommand{\la}{\langle} 

\newcommand{\ra}{\rangle}

\newcommand{\rp}{r_\perp} %\newcommand{\non}{\nonumber\\}

\def\jmo#1{{ J.\ Mod.\ Opt.} {\bf#1}}

\def\ol#1{{ Opt.\ Lett.} {\bf#1}}
\def\pla#1{{ Phys.\ Lett. A\/} {\bf#1}}

\def\pra#1{{ Phys.\ Rev. A\/} {\bf#1}}

\def\prl#1{{ Phys.\ Rev.\ Lett.} {\bf#1}}

\begin{document}

%\title{ Polarization Coherence Theorem v7b}
\title{Polarization Coherence Theorem}
\author{J.H. Eberly, X.-F. Qian and  A.N. Vamivakas}
\address{Center for Coherence and Quantum Optics,\\
The Institute of Optics and Department of Physics \& Astronomy, University of Rochester, Rochester, NY 14627, USA}

\begin{abstract}
Visibility $V$ and distinguishability $D$ quantify wave-ray duality: $V^2 + D^2 \le 1$. We join them to polarization $P$ via the Polarization Coherence Theorem, a tight equality: $P^2 = V^2 + D^2$.\\

\end{abstract}

\maketitle

Centuries after Thomas Young's famous double-slit interference experiment \cite{Young}, quantification of coherence as a contextual resource (see \cite{Eberly-etal}) is just now being examined (see \cite{Plenio, Adesso}), and experimental evidence of polarization coherence in a previously unexplored context has been reported, exposing a new coherence triad \cite{QMVE}. 

The theorem of the title arises from the recognition that polarization is a two-party property, even in common usage. A polarized electorate means two things: (a) opinions within the population favor only a few political factions and (b) this occurs within a specified background (the voting public, but not kindergarten school children). In much the same way in optics a polarized field means that just one or two of the field's intrinsic spin orientations are endowed with substantial field amplitude, so two independent degrees of freedom, spin and amplitude, are well correlated.  Here we show how this leads to the new identity we refer to as the Polarization Coherence Theorem (PCT). 

\begin{figure}[b!]
\includegraphics[width=5cm]{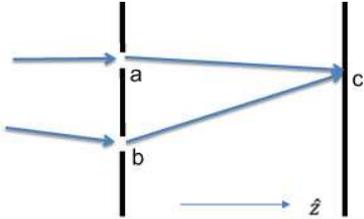}
\caption{Double-slit interference setup. Classical light fields emerge from slits $a$ and $b$ and combine on screen $c$.} 
\label{doubleslit}
\end{figure}

For simplicity we introduce the PCT in the most familiar context, Young's double-slit scenario as shown in Fig.~\ref{doubleslit}. The combined amplitude of the light field arriving at the screen has a contribution from each of the two slits $a$ and $b$:  
\beq \label{F-field}
F(\rp,z) = u_a(\rp,z)\Phi_a(q) + u_b(\rp,z) \Phi_b(q). 
\eeq
Here $u_a(\rp,z)$ and $u_b(\rp,z)$ are diffractive spatial mode functions unit-normalized and orthogonal in the intervening space from the slits to the screen, and $\Phi_a$ and $\Phi_b$ are the corresponding field strengths. They depend on degrees of freedom not identified and here loosely labeled $q$, such as temporal amplitude, spin (ordinary polarization), etc. 

By adopting the conventional small-angle and distant-screen treatment of the Young signals, the propagation factors from slits $a$ and $b$ to screen $c$ will be the same in magnitude but differ in phase (which is absorbed into amplitudes $\Phi_a$ and $\Phi_b$). Then the intensity at the screen is obtained with the expected sinusoidal interference term:
\beq
I_c = I_{a} + I_{b} + 2| \la\Phi_a^*\Phi_b\ra|\cos[\arg(\la\Phi_a^*\Phi_b\ra)], \label{I_c-perfect}
\eeq
with $I_{a} = \la|\Phi_a|^2\ra$ and $I_{b} = \la|\Phi_b|^2\ra$, and the brackets for averaging are needed if the amplitudes are known only statistically, as is commonly the case.

Fringe visibility for field $F$ is given as
\beq \label{V}
V_F = \frac{I_c^{\max} - I_c^{\min}}{I_c^{\max} + I_c^{\min}} = \frac{2|\langle \Phi^{*}_a \Phi_b \rangle| }{I_{a} + I_{b}}, 
\eeq
where $I_c^{\max}$ and $I_c^{\min}$ are the maximum and minimum fringe intensities of $I_c$ as the cosine equals $+$1 or $-$1 in Eqn.~(\ref{I_c-perfect}). 

When light is coming from only one of the two slits then there can be no interference and the origin of the light arriving at screen $c$ is known. Its intensity is distinguishable from the intensity observed when only the other slit is open. Consequently, the degree of distinguishability is also an indicator of a different  coherence, and is defined as 
\beq \label{D}
D_F = \frac{|I_{a} - I_{b}|}{I_{a} + I_{b}}. 
\eeq
One can easily check (Schwarz inequality) that $\la|\Phi_a|^2\ra\la|\Phi_b|^2\ra \ge |\langle \Phi^{*}_a \Phi_b \rangle|^2$, and conclude that $V_F$ and $D_F$ satisfy a strong constraint: 
\beq \label{V2D2ineq}
V_F^2 + D_F^2 \le 1. 
\eeq
This coherence relation, here obtained without considering spin (ordinary polarization) at all, when interpreted in terms of wave/ray or wave/particle duality, occupies a central place in all discussions of duality and complementarity, and has been derived and re-derived by many investigators  \cite{V2D2proofs}.  

The Young scenario can also be discussed for an optical field exhibiting polarization unit vectors explicitly:
\beq \label{E-field}
E = \hat h E_h + \hat v E_v.
\eeq
Classical polarization optics tells us (see, e.g., Brosseau \cite{Brosseau} or Wolf \cite{Wolf 2007}) how to construct ${\cal W}_E$, the polarization coherence matrix for $E$ and its degree of  polarization $P_E$, indexed by spin orientations $\hat h$ and $\hat v$. They are:
\beq \label{Ematrix}
{\cal W}_E  = \left[\begin{matrix}
\la E_h^* E_h\ra  & \la E_v^* E_h\ra  \\
\la E_h^* E_v\ra   & \la E_v^* E_v\ra  \\
\end{matrix} \right],\ P_E =\sqrt{1 -\frac{4 Det {\cal W}_E}{(Tr[{\cal W}_E])^2}}.
\eeq 

Note that polarization also applies to field $F$, despite the absence of directional unit vectors. We can say that $F$ is strongly ``polarized" in the $u_a$ ``direction" if $|\Phi_a(q)|$ has much greater magnitude than $|\Phi_b(q)|$, and the reverse if $|\Phi_b(q)| \gg |\Phi_a(q)|$. There is no intrinsic mathematical difference in the $E$-polarization and $F$-``polarization" examples, since $\hat h$ and $\hat v$, and the $u_a$ and $u_b$ functions, are orthogonal vectors in their own vector spaces of spin and spatial mode. We will designate the $u_a$ and $u_b$ mode functions as ``mode polarization vectors". The field in (\ref{F-field}) exhibits ``mode polarization" in its mode space.

Thus we easily construct ${\cal W}_F$, the ``mode polarization" coherence matrix for $F$ in (\ref{F-field}), and its ``degree of mode polarization" $P_F$, indexed by ``mode polarization" orientations $u_a$ and $u_b$. They are:
\beq \label{Fmatrix}
{\cal W}_F  = \left[\begin{matrix}
\la \Phi_a^* \Phi_a\ra  & \la \Phi_b^* \Phi_a \ra \\
\la \Phi_a^* \Phi_b\ra  & \la \Phi_b^* \Phi_b\ra \\
\end{matrix} \right], \quad
  P_F =\sqrt{1 -\frac{4 Det {\cal W}_F}{(Tr[{\cal W}_F])^2}}.
\eeq 
Elementary arithmetic then provides:
\beqa \label{modeP}
P_F^2 &=& 1 - 4\frac{I_aI_b - |\la \Phi_a^* \Phi_b\ra|^2}{(I_a+I_b)^2}\nonumber \\
&=& \frac{(I_a - I_b)^2 + 4|\la \Phi_a^* \Phi_b\ra|^2}{(I_a+I_b)^2} \nonumber \\
&=& \frac{(I_a - I_b)^2}{(I_a+I_b)^2} + 4\frac{|\la \Phi_a^* \Phi_b\ra|^2}{(I_a+I_b)^2},\nonumber \\
\eeqa
and this contains elements familiar from (\ref{V}) and (\ref{D}). In fact, we have just derived
\beq \label{P2V2D2}
P_F^2 \equiv D_F^2 + V_F^2. 
\eeq
The $F$ degree of polarization satisfies $ 1 \ge P_F \ge 0$, so there is no conflict between the result (\ref{P2V2D2}) and the multiply derived formula (\ref{V2D2ineq}). 

It should be obvious that the new identity (\ref{P2V2D2}) arises equally quickly from (\ref{Ematrix}) as from (\ref{Fmatrix}). That is, it's independent of the pair of degrees of freedom used to derive it. The identity (\ref{P2V2D2}) is in fact a very general Polarization Coherence Theorem. It says that the degree of polarization coherence for an optical field, in either the ordinary spin sense or a generalized mode sense, will support the famous duality inequality (\ref{V2D2ineq}). As such, (\ref{P2V2D2}) has unexpected implications for complementarity, which will have to be discussed elsewhere \cite{complem}, as well as the single-photon version of the present result.\\

\noindent {\em Appendix} We are using analogies that have a firm relationship because they share the exact same vector space basis, and there is a clear connection between the two fields (\ref{F-field}) and (\ref{E-field}) and thus between our two ``polarizations". Recall that the amplitudes $\Phi(q)$ are dependent on additional degrees of freedom of the light field, unspecified but labeled $q$ above. Similarly, the amplitudes $E_h$ and $E_v$ depend on spatial and temporal mode amplitudes, not specified but in the relevant Young context they certainly refer to contributions coming from the spatial modes $u_a$ and $u_b$, so we can expand $E_h$ and $E_v$ in terms of them:
\beq \label{Eab}
E_h = u_a H_a + u_b H_b,\ {\rm and}\ E_v = u_a V_a + u_b V_b.
\eeq
Then the otherwise arbitrary field $F$ can be interpreted as just the projection of $E$ on an arbitrary intrinsic spin direction $\hat s$:
\beqa
F &=& (\hat s \cdot E) = (\hat s \cdot \hat h) (u_a H_a + u_b H_b) \nonumber \\
&+& (\hat s \cdot \hat v) (u_a V_a + u_b V_b) \nonumber \\
&=& u_a [(\hat s \cdot \hat h)H_a + (\hat s \cdot \hat v) V_a ] + u_b [(\hat s \cdot \hat h)H_b + (\hat s \cdot \hat v) V_b ], \nonumber
\eeqa
which is the same as $F =  u_a \Phi_a  + u_b \Phi_b$.\\ 

\noindent Funding: University of Rochester Research Award, Leonard Mandel Faculty Fellowship in Quantum Optics, ARO W911NF-16-1-0162, ONR N00014-14-1-0260, NSF (PHY-1203931, PHY-1505189, PHY-1539859).\\

\end{document}